\documentclass[useAMS,usenatbib,usegraphicx]{mn2e}

\title[On the accuracy of the high-z cluster luminosity function]{On the accuracy of the high redshift cluster luminosity function }
\author[R. P. Mu\~noz, N. D. Padilla, and L.F. Barrientos]{R. P. Mu\~noz$^{1}$\thanks{E-mail:
rmunoz@astro.puc.cl}, N. D. Padilla$^{1}$ and L.F. Barrientos$^{1}$\\§
$^{1}$Universidad Cat\'olica de Chile, Avda. Vicu\~na Mackenna 4860, Casilla 306, Santiago 22, Chile.}
\begin{document}

\date{}

\pagerange{\pageref{firstpage}--\pageref{lastpage}} \pubyear{2008}

\maketitle

\label{firstpage}

\begin{abstract}
We study the reliability of the statistical background subtraction method for computing 
the Ks-band luminosity function of cluster galaxies at $z\sim1$ using
mock Red-sequence Cluster Survey cluster catalogues constructed from GALFORM semi-analytic galaxies. 
The underlying cluster luminosity function in the mocks are compatible with recent
estimates at $z\sim 1$ by several authors.
 We simulate
different samples where the number of clusters with Ks-band photometry goes from $5$ to
a maximum of $50$, in order to find the most suitable observational sample to carry out
this study; the current observational status in the nIR wavelength range has been reached using 
$5$ real clusters at $z\sim1$.
We compute the composite luminosity function 
for several samples of galaxy clusters with masses  $\simeq 1.5\times10^{14} M_{\odot}$ 
assuming a flux limited,
complete sample of galaxies down to ${\rm Ks}=21.0$ magnitudes.  We find that the Schechter 
fit parameters $\mathrm{Ks}^*$ and $\alpha$ for a sample of galaxies with no redshift information 
are rather poorly constrained if both parameters are allowed to vary freely;
if $\alpha$ is fixed at a fiducial value, then $\mathrm{Ks}^*$ shows significantly
improved stochastic uncertainties but can be influenced
by systematic deviations.
We find a significantly improved accuracy in the luminosity function parameters 
when adding photometric redshift information for bright cluster galaxies.
The impact of a ten-fold increase in the
number of clusters with available Ks-band photometry is that of decreasing
stochastic errors in $\mathrm{Ks}^*$ and $\alpha$ by 
factors of $\simeq 2$ and $\simeq 4$, respectively, for 
accuracies of up to $\left(\Delta \mathrm{Ks}^*/\mathrm{Ks}^*\right)_{\rm stochastic}=
0.02$ and $\left(\Delta \alpha/\alpha\right)_{\rm stochastic}=0.09$.
The dwarf-to-giant ratios inferred from the luminosity functions of red-sequence galaxies in the mock catalogue 
agree very well with
the underlying values; however, there is an indication that the semi-analytic model 
over-predicts the abundance
of dwarf galaxies by up to a factor of $3$ with respect to recent measurements. Finally, we find that 
in order to use estimates of ${\rm Ks}^*$
to study the formation redshift of cluster galaxies at $z=1$, the sample would need to contain
520 $z\sim1$ clusters, for
an accuracy of $\sim 2\; \mathrm{Gyr}$ at the 68 per cent confidence level.  However,
combining this method with other estimates may reduce significantly the sample size, and allow important new
constraints on galaxy formation models.
\end{abstract}

\begin{keywords}
galaxies:clusters -- galaxies:evolution -- galaxies:luminosity function
\end{keywords}

\section{Introduction}

The galaxy luminosity function (LF) is one of the fundamental quantities of observational cosmology.
Its evolution with redshift and dependence on galaxy morphology and environment
has been extensively used to provide strong constraints on galaxy evolution models.

Clusters of galaxies are the largest virialised structures at any given epoch, which grow from 
high-density regions in the primordial matter distribution of the Universe, and may have formed
between $9\;\mathrm{Gyrs}$ ago and the present \citep{sta06}. 
Most of them have been detected in surveys carried out in the X-ray (\citealt{rom01}, \citealt{bar06}) and optical bands (\citealt{pos96}, \citealt{gon01}, \citealt*{gla05}), and it is expected that upcoming Sunyaev-Zel'dovich surveys will yield thousands of galaxy clusters \citep{car02}.
The fact that clusters contain large numbers of galaxies
practically at the same distance from the observer within a small area on the sky,
makes them ideal systems to study the LF down to very faint magnitudes.

In the local Universe, the LF of cluster galaxies has been studied extensively by \citet{got02} and \citet{dep03}, 
who found that there is no significant evidence for variations in the LF across a broad range of cluster properties, and also, that
the cluster LF differs significantly from the field LF.
Although many efforts have been done to perform such a study at $z\sim1$, a significant look-back 
time of $7\;\mathrm{Gyrs}$ in the currently favoured cosmology,
only a handful of high redshift clusters have been detected to date in optical and X-ray surveys (\citealt{sta97}, \citealt{tof04}, \citealt{dep07}). Recently, \citet{eis08} found 106 galaxy clusters and groups candidates at $z>1$ in the Spitzer Infrared Camera (IRAC) Shallow Survey \citep{eis04}.

Obtaining spectroscopic measurements for a large number of
faint cluster galaxies at $z\sim1$ is a very difficult 
and time-expensive task.  As a consequence, 
to date there are no reliable LFs measured at $z\sim1$ using only spectroscopically 
confirmed members. Several authors have attempted to solve 
this issue by using photometric redshifts to define cluster membership 
and to compute the LF (\citealt{tof04}, \citealt{tan07}), but their results have been criticised due to
the unknown spectral energy distribution (SED) of galaxies at high redshifts 
(\citealt{and05}, {\citealt{she07}).

Traditionally, the cluster LF is computed applying a background subtraction method, which consists of computing 
the difference between galaxy counts in the cluster and control field directions.  Several authors have developed and 
applied their own subtraction methods (\citealt{oem74}, \citealt{pim02}, \citealt{and05}).  However, it is difficult
to assess whether this method is free of important systematic effects due to possible problems in the background
subtraction method, which needs to select an appropriate area, either around the clusters under study or from
an independent large area photometric sample.

It has been claimed that infrared luminosities are better suited to measure the LF than their optical counterparts, 
since the former are comparatively insensitive to the star formation history while reflecting the total stellar mass
(\citealt{cha96}, \citealt{mad04}). 
Furthermore, infrared luminosities have a negligible extinction correction and their K-correction depends 
only weakly on Hubble type (see for instance \citealt{man01}).

The main goal of this work is to study the reliability of the statistical background subtraction method 
to recover the underlying observer-frame Ks-band LF. Section 2 explains the mock catalogues in which we base our analysis. 
Section 3 discusses the background subtraction method and shows the resulting LF estimates, and
Section 4 shows a comparison between the underlying and recovered LF, and several
physical quantities inferred from the LF.  Our conclusions are summarized in Section 5.

Throughout this paper we assume $\Omega_M = 0.3$, $\Omega_\Lambda = 0.7$ and $H_0 = 70~km \,s^{-1}\,Mpc^{-1}$. 
All magnitudes are in the Vega system.

\section{Data}

This work concentrates in particular, on the possibility to study the LF of $z\simeq1$ clusters, in such
a way so as to obtain the accuracy needed to study different properties of the galaxy population
including the relative importance of dwarf vs. giant galaxies (\citealt{del07}, \citealt*{gil08a}), or the
expected formation redshift of $z\sim1$ cluster galaxies.  To date,
the most suited cluster survey for such a study is the Red sequence Cluster Survey (RCS-1, \citealt{gla05})
However, there is only optical information for these clusters, so a study of the LF over the infrared
wavelength range would require an observing campaign to obtain these data.  The authors of this work 
are at the moment in the process of obtaining such information, so that this study will be possible in 
the near future.

\subsection{Mock catalogues}

In order to test for possible systematic errors affecting a statistical measurement of the
cluster LF, we apply this analysis to mock RCS catalogues.  
We construct mock RCS catalogues using the Millennium simulation \citep{spr06}
populated with semi-analytic galaxies from the GALFORM model by \citet{bau05} (see also 
\citealt{col00}, \citealt{bow06}; for an alternative approach see \citealt{lag08}).  The numerical simulation 
follows the gravitational interaction of $10^{9}$ dark-matter particles 
in a $\Lambda$CDM Universe characterized by the cosmological parameters $\Omega_m=0.30$, 
$\Omega_{\Lambda}=0.70$,
a power law spectral index $n=0.94$, an amplitude of matter density fluctuations in spheres 
of $8\; \mathrm{h}^{-1}\, \mathrm{Mpc}$,
$\sigma_8=0.8$, and a Hubble constant $H_0=100\; h\, \mathrm{km}\, \mathrm{s}^{-1}\, \mathrm{Mpc}^{-1}$ with $h=0.70$.  The GALFORM semianalytic model
was applied to the merger trees of halos in the simulation, which have a minimum mass of $1.2\times10^{10}\; \mathrm{h}^{-1}\, \mathrm{M}_{\odot}$
for a minimum of 20 dark-matter particles (courtesy of the Durham group).  
The galaxies in the simulation are characterized by their 
luminosity in several photometric bands, including the near infrared Ks-band and those 
available in the RCS dataset.  In order to construct each mock catalogue, we place
an observer in the
simulation, and record all angular positions of galaxies down to the RCS magnitude limit within an area of
$100$ square degrees.  
The intrinsic completeness limit in our mock catalogues is $\mathrm{Ks}=22.0$, observer-frame, 
following the resolution constraints of the semi-analytic galaxy population.  Notice that
this limit can be lowered to meet the restrictions of a given observational sample.

In order to include the evolution of the galaxy population with redshift, we use simulation
outputs corresponding to four different redshifts from $z=0.3$ to $z=3$ for consecutive redshift
ranges. This ensures that the  population of background galaxies behind our mock RCS clusters
are characterized by an evolving luminosity function to distances beyond the maximum expected galaxy
redshifts allowed by the magnitude limit cut.
The fact that we are not interested in the differential evolution
of the luminosity function with redshift justifies our choice of mock light-cones with discretised evolution.

The advantage of using mock catalogues relies in that the underlying population of clusters is known, along with
the LF parameters for the field and cluster environments for each individual
mock.  Furthermore, as we are using multiple mocks extracted from practically independent volumes in the
simulation, we can estimate the expected effects of sample variance in our statistics.  
Besides these advantages, we are also
able to make a comparison between the properties of galaxies populating the mocks and those
in real datasets published in previous works, providing a new test for the GALFORM semi-analytic model.

Since we are interested in employing these mock catalogues to study the accuracy of the background subtraction method
for computing the galaxy LF for RCS-1 clusters, we compare the mass function between simulated and observed clusters.
Figure \ref{mass_function} shows the mass function of simulated clusters selected from the mock catalogues and RCS-1 observed clusters
taken from \citet{gla07}, where it can be seen that both mass functions are roughly compatible for $M_{200} > 1.5\times 10^{14}\, M_\odot\, h^{-1}$ to a $1\,\sigma$ level. The offset in the observational data can be due in part to errors in the mass measurements which usually broaden
statistical distributions (see for instance \citet*{pad99})

\begin{figure}
\includegraphics[width=8cm]{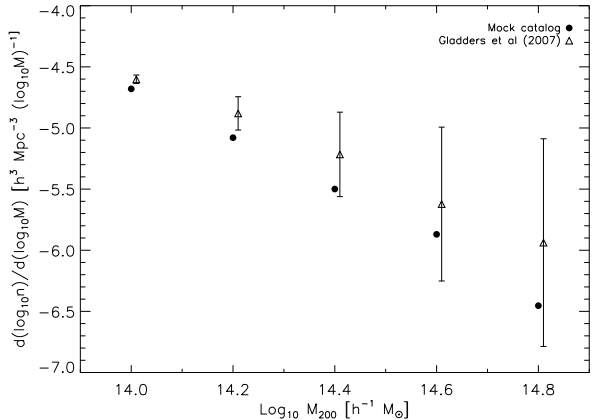}
\caption{Mass function of simulated and observed clusters. The filled circles show the mass function of simulated clusters selected
from the mock catalog. The open triangles show the mass function of observed clusters selected from the RCS-1 cluster catalog \citep{gla07}.}
\label{mass_function}
\end{figure}

\section{Luminosity function estimates}

The optimal method to build a reliable cluster LF consists in using spectroscopic information to 
determine cluster membership.
However, obtaining spectroscopic redshifts for a large number of faint galaxies at 
high redshifts is prohibitively expensive in
telescope time. 

The cluster samples in our mock catalogues are characterized by a $z=1$ LF in the Ks-band which
we fit with a Schechter function of characteristic luminosity $\mathrm{Ks}^*=18.27$ and faint-end slope $\alpha=-1.36$.
Our approach to determine the reliability of different LF estimators takes advantage of our knowledge of the underlying LF parameters. 
We notice that the characteristic luminosity in the model is in good agreement
with the observational result $\mathrm{Ks}^*=18.5\pm1.2$ found by \citet{str06}, who applied a background subtraction 
method and fitted for both, the characteristic luminosity ${\rm Ks^*}$ and $\alpha$, the faint-end slope. 
Other measurements show different levels of agreement; for instance, \citet{kod03} found a fainter value 
$\mathrm{Ks}^*=17.57\pm0.24$, using a fixed value of $\alpha=-0.9$ and relatively shallow nIR data.
With respect to the faint-end slope, this parameter is poorly constrained by observational data, and consistent with
$\alpha=-1$.  Therefore, down to the accuracy of the observational results, the semi-analytic model used to construct
the mock catalogues is able to reproduce the observed cluster galaxy population at $z\simeq 1$.

In the following subsection we discuss two possible LF estimators that we will apply to the mock cluster catalogues.

\subsection{Background subtraction and photometric redshift LF estimators}

Two widely used methods to calculate the LF
are the statistical background subtraction and the photometric redshift method. 
The former consists on using one or several control fields to determine the number of contaminating galaxies (background 
and foreground galaxies) per unit area as a function of magnitude, to then compute the galaxy LF as the difference 
between the galaxy number count in the cluster direction and the control field. 
This is the method of choice when there is photometry in only one or two bands for galaxies in the cluster direction,
and very few cluster members have been spectroscopically confirmed. Employing numerical simulations, it has been shown that the background subtraction method can recover accurately the underlying LF of clusters selected in three dimensions \citep{val01}.

The background subtraction method requires to define a magnitude bin width, a cluster-centered aperture, and a control 
field region.  When computing the number counts in the cluster direction for the RCS mock catalogue, we 
mimic the observational method of employing galaxies enclosed by 
a circular aperture of radius $0.27\; \mathrm{h}^{-1}\; \mathrm{Mpc}$ centered on the brightest cluster galaxy and a magnitude bin of width $0.5\; \mathrm{mag}$.
This aperture size is justified because the nIR imaging of RCS clusters covers only $0.8\; \mathrm{h}^{-1}\; \mathrm{Mpc}$ of the cluster
central region. We defined the control field region based in the observational nIR surveys to date, which should have a minimum $5\sigma$
limiting magnitude of $\mathrm{Ks}=22.0$ and cover a big area in the sky. The candidates were the Faint Infrared Extragalactic Survey
(FIRES; \citealt{lab03}), which covers $8.3\; \mathrm{arcmin}^2$ and reaches a $5\sigma$ depth for point sources of
$\mathrm{Ks_{5\sigma}}=24.5$; the Great Observatories Origins Deep Survey of CDF-S region
(GOODS/ISAAC; \citealt{ret08}), which covers $170\; \mathrm{arcmin}^2$ and reaches $\mathrm{Ks_{5\sigma}}=22.8$;
and the nIR VIMOS-VLT Deep Survey (VVDS-NIR; \citealt{iov05}),
which covers $430\; \mathrm{arcmin}^2$ and reaches $\mathrm{Ks_{5\sigma}}=22.1$.
Figure \ref{control_field_comparison} shows the ratio between the rms variation of number count along the control field regions previously mentioned and the rms variation of actual cluster members in the mock. We can see that a control field of size
the FIRES survey contributes about $30\%$ to the rms variation along the cluster direction, so the accuracy of the LF built from a background
subtraction method would be highly limited by the small control field region. Also, in order to reach a $0.4\%$
contribution of background rms variation to the Ks-band LF error bars we would need a nIR survey covering $1\; \mathrm{deg}^2$, which does
not even exist at this magnitude depth to date. This way, GOODS/ISAAC and VVDS-NIR are the most suitable surveys to build the Ks-band LF
with about $100\%$ completeness to magnitude 21.0 using the background subtraction method.

We discard the VVDS-NIR survey, since the calibrated images of GOODS are publicly available and are taken with the same instrument
as the nIR imaging of the RCS clusters \citep{mun08}

Using the mock catalogues, we compute the rms variation of number
counts from  several $170\; \mathrm{arcmin}^2$ regions in the mock catalogues, and compare this to Poisson errors and jacknife algorithm estimates (see Figure \ref{clf_error_comparison}, top panel).
It can be seen that a Poisson distribution underestimates the actual rms variation
about $30\%$ and $60\%$ for bright and faint galaxies, respectively. This high discrepancy can be explained by the galaxy-galaxy correlation
and the presence of large scale structure along the control field region (see for instance, \citealt{pad04}, \citealt{paz08}). We also see that the
jacknife algorithm reproduces the underlying spread in number counts seen in the simulation to a higher degree. This way, the error
in the galaxy count along the control field region, $\sigma_{N_{f}}$, will be computed from the jacknife algorithm.

\begin{figure}
\includegraphics[width=8cm]{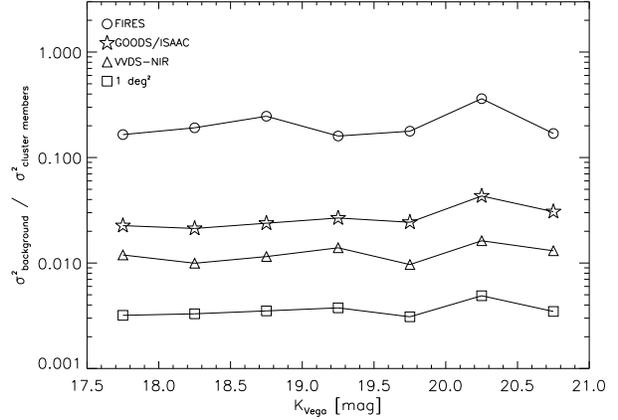}
\caption{ Ratio between the rms variation of number counts along the control field regions of several sizes and the rms variation of
actual cluster members.
Open circles correspond to the deepest nIR survey FIRES, stars correspond to the public ESO survey of the CDF-S region GOODS/ISAAC,
and triangles to the deep and wide VVDS. The squares correspond to an hypothetical nIR survey covering $1\; \mathrm{deg}^2$,
reaching a $5\sigma$ limiting magnitude deeper than $\mathrm{Ks}=22.0$}
\label{control_field_comparison}
\end{figure}

The observational program being carried out to obtain near-infrared imaging of RCS clusters
(\citealt{mun08}) will only cover fields centered in clusters, therefore, it is important to test whether the use of number counts
from a different source catalogue will still allow measuring the LF with minimum systematic deviations from the
underlying values.

The number of clusters members along the cluster direction in the $j^{th}$ magnitude bin $N_{j}$, and its error, $\sigma_{N_{j}}$, are
given by
 \begin{eqnarray}
N_{j} & = & N_{dj} - N_{fj} \frac{\Omega _d}{\Omega _f} \\
\sigma_{N_j} & = & \sqrt{ \sigma_{N_{dj}}^2 + \left( \sigma_{N_{fj}} \frac{\Omega _d}{\Omega _f} \right)^2 },
\end{eqnarray}
where $N_{dj}$ is the number of galaxies (cluster+field) in the cluster direction in the $j^{th}$ bin, $N_{fj}$ 
is the number of galaxies in the control field region in the $j^{th}$ bin, $\Omega _d$ is the area of the 
cluster-centered circular aperture, and $\Omega _f$ is the area of the control field region. The error in 
the number counts in the cluster direction, $\sigma_{N_{dj}}$, is
estimated by the Poissonian error, a choice we justify next.

\begin{figure}
\includegraphics[width=8cm]{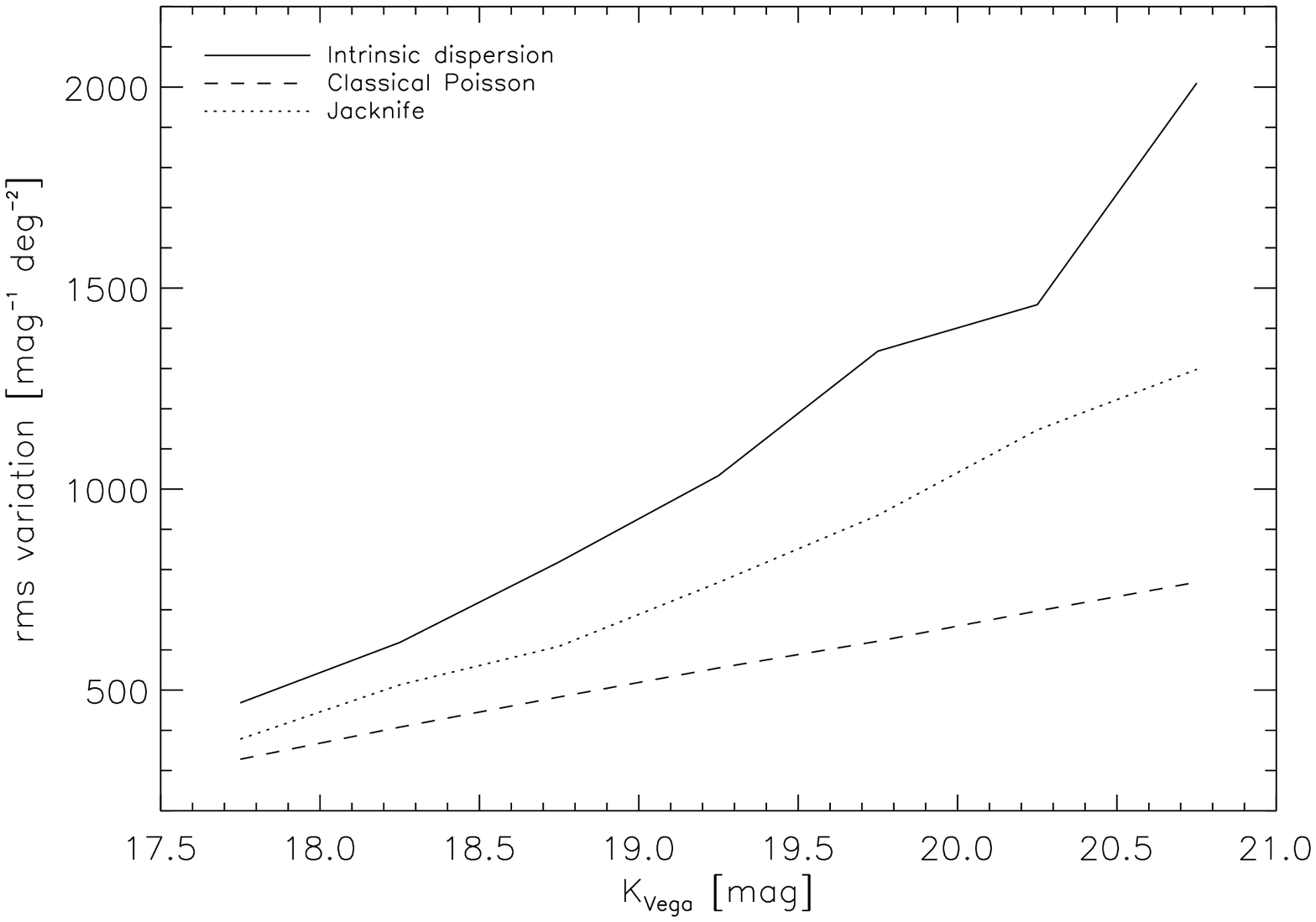}
\includegraphics[width=8cm]{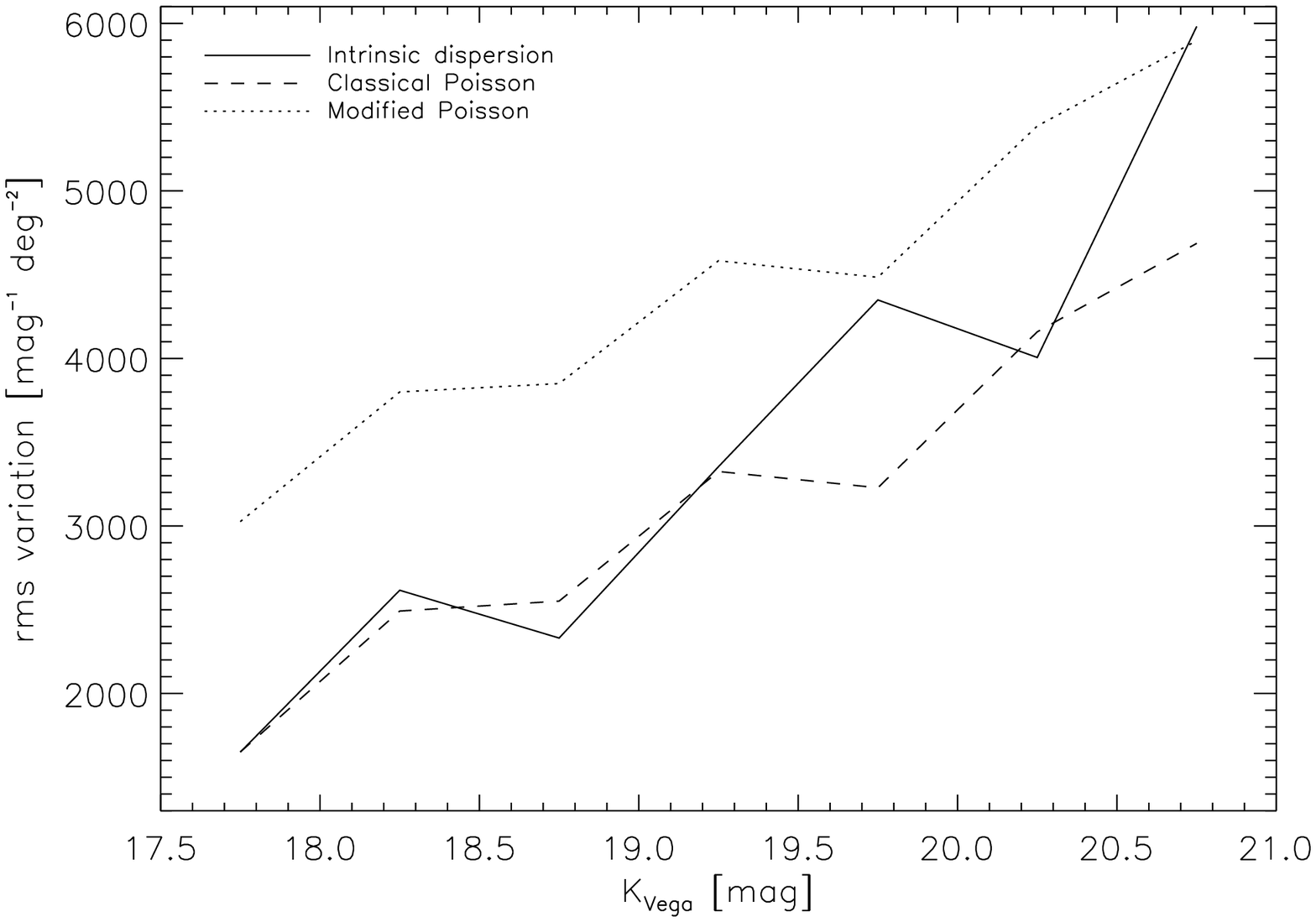}
\caption{ rms variation in the galaxy counts along the control field region of size the GOODS/ISAAC survey (top panel) and along the 
cluster direction (bottom panel) as function of magnitude. 
The solid line shows the actual dispersions built from $100$ independent mock samples. 
The dashed and dotted lines in the top panel show the errors in the background number count estimated from the classical Poissonian
error $\sqrt{N}$ and from applying a jacknife technique, respectively. The dashed and dotted lines in the bottom panel show the errors in
cluster direction number count estimated from the classical and a modified Poissonian error $1+\sqrt{N+0.75}$ }
\label{clf_error_comparison}
\end{figure}

We calculate the counts in the cluster direction for $100$ independent samples containing
$10$ RCS clusters each, and calculate the dispersion in these counts as a function of
apparent magnitude.  We then select
one of these samples at random, which we use to
compute the classical Poisson error $\sqrt{N}$ for the number counts
in the cluster direction, and the modified Poisson error $1+\sqrt{ (N_{dj}+0.75)}$ proposed 
by \citet{geh81} and employed by several authors (\citealt{and05}, \citealt{str06}).
The bottom panel of Figure \ref{clf_error_comparison} shows the comparison between the 
dispersion in the counts in the cluster direction and its estimated error as a function of the observed magnitude. From 
this figure we conclude that the classical Poisson error resembles best the dispersion in the mocks, 
while the modified Poisson errors overestimates errors by a significant amount.  Therefore,
this justifies our choice of Poisson errors when studying the cluster counts in the mock CLFs.  
\citet{zhe05} find that the number of galaxies in dark matter haloes in semi-analytic and SPH simulations 
follows a Poisson distribution,
which would indicate 
that galaxies outside the cluster (either in front or behind due to projection effects)
only contribute a relatively low fraction of the uncertainty in the cluster counts.

\begin{figure}
\includegraphics[width=8cm]{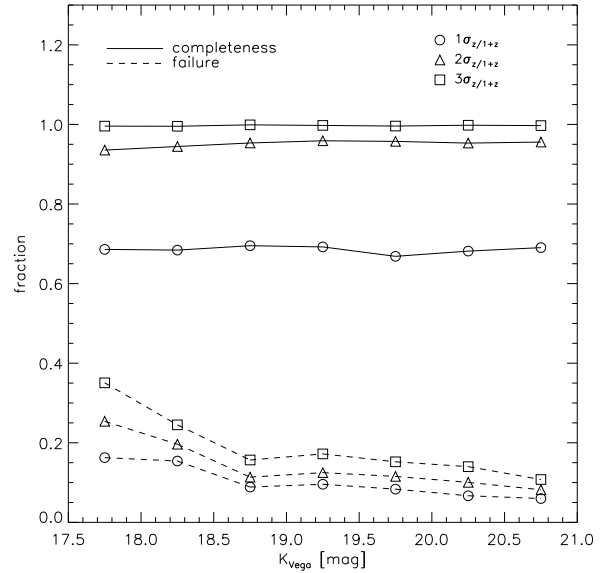}
\caption{ Completeness (\textit{solid line}) and failure ratio (\textit{dashed line}) of photometric redshift method as function of Ks-band magnitude.
The open circles correspond to cluster members candidates lying 1 $\sigma_{z / (1+z)}$ from the cluster redshift, while the triangles and squares
correspond to galaxies lying 2 and 3 $\sigma_{z / (1+z)}$, respectively.}
\label{photoz_completeness}
\end{figure}

The photometric redshift method consists of using broad-band photometry in several band-passes as a very low resolution spectrum,
and then fitting a spectral energy distribution (SED) template to estimate the redshift of the galaxies in the cluster field.  
The final step consists on defining as cluster members those galaxies lying within a redshift shell centered on the cluster redshift.
\textbf{ For simplicity , we do not compute the photometric redshifts of galaxies from fitting a SED to the observed magnitudes but rather
from perturbing their actual redshifts by a gaussian distribution with $\sigma_{z / (1+z)} = 0.05\,$. It is important to note
that the true probability distributions of photometric redshifts are often highly non-Gaussian, and that secondary maxima or extended wings might
change the accuracy of photometric redshift method used in this paper.}
Figure \ref{photoz_completeness} shows the completeness and failure ratio of the photometric redshift method computed from 500 independent
Monte-Carlo realizations of photometric redshift of randomly selected clusters. 
\textbf{Each realization consists in choosing a random cluster located at $z\sim1$, then selecting
those galaxies within a circular aperture of radius $0.27\; \mathrm{h}^{-1}\; \mathrm{Mpc}$, and finally, simulate their
photometric redshifts using the actual redshifts from the mock catalogues via a Monte-Carlo procedure.}
We can see that the photometric redshift cut
$\textrm{1-}\sigma_{z / (1+z)}$ recovers a low percentage of actual cluster galaxies and has a failure ratio of about 10 percent, while the redshift cut of
$\textrm{2-}\sigma_{z / (1+z)}$ recovers almost all the actual cluster galaxies and has a failure ratio of about 15 percent. The cut of
$\textrm{3-}\sigma_{z / (1+z)}$ has the highest completeness, but the failure ratio of bright galaxies increases rapidly. According to the previous considerations, we define
as cluster members those galaxies lying within $\textrm{2-}\sigma_{z / (1+z)}$ from the cluster redshift. Large scale structure plays a role in this calculations and therefore a simple analytic gaussian is only approximate.

The RCS clusters under study by \citet{mun08} have deep J and Ks-band imaging, and shallow $\rm R_C$ and z imaging, so it  is not possible
to build the LF up to magnitude $Ks=21.0$ from applying the photometric redshift method. In order to take advantage of the available data,
we propose a new method that consists in using photometric redshift of galaxies to compute the bright part of the LF and using the background
subtraction method to compute the faint part. The cluster members with $\rm{Ks} \leq 18.5$ are identified by their photometric redshift,
while the number of members with $\rm{Ks} > 18.5$, by using the background subtraction method. The combination of both methods will be called
B+Z method hereafter.


\begin{figure*}
\centering
\begin{minipage}[b]{0.5\textwidth}
\centering
\includegraphics[width=8cm]{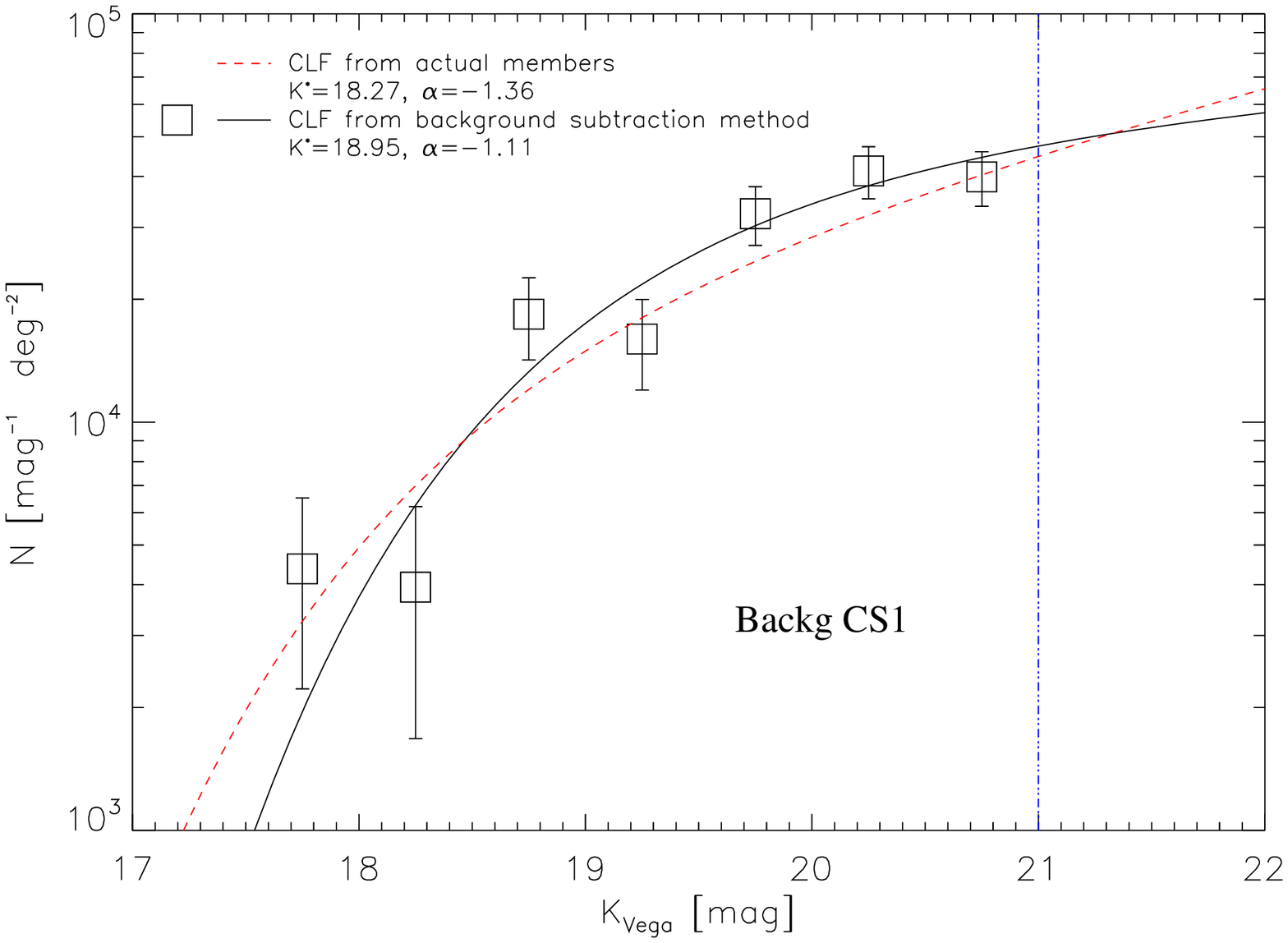}
\end{minipage}%
\begin{minipage}[b]{0.5\textwidth}
\centering
\includegraphics[width=8cm]{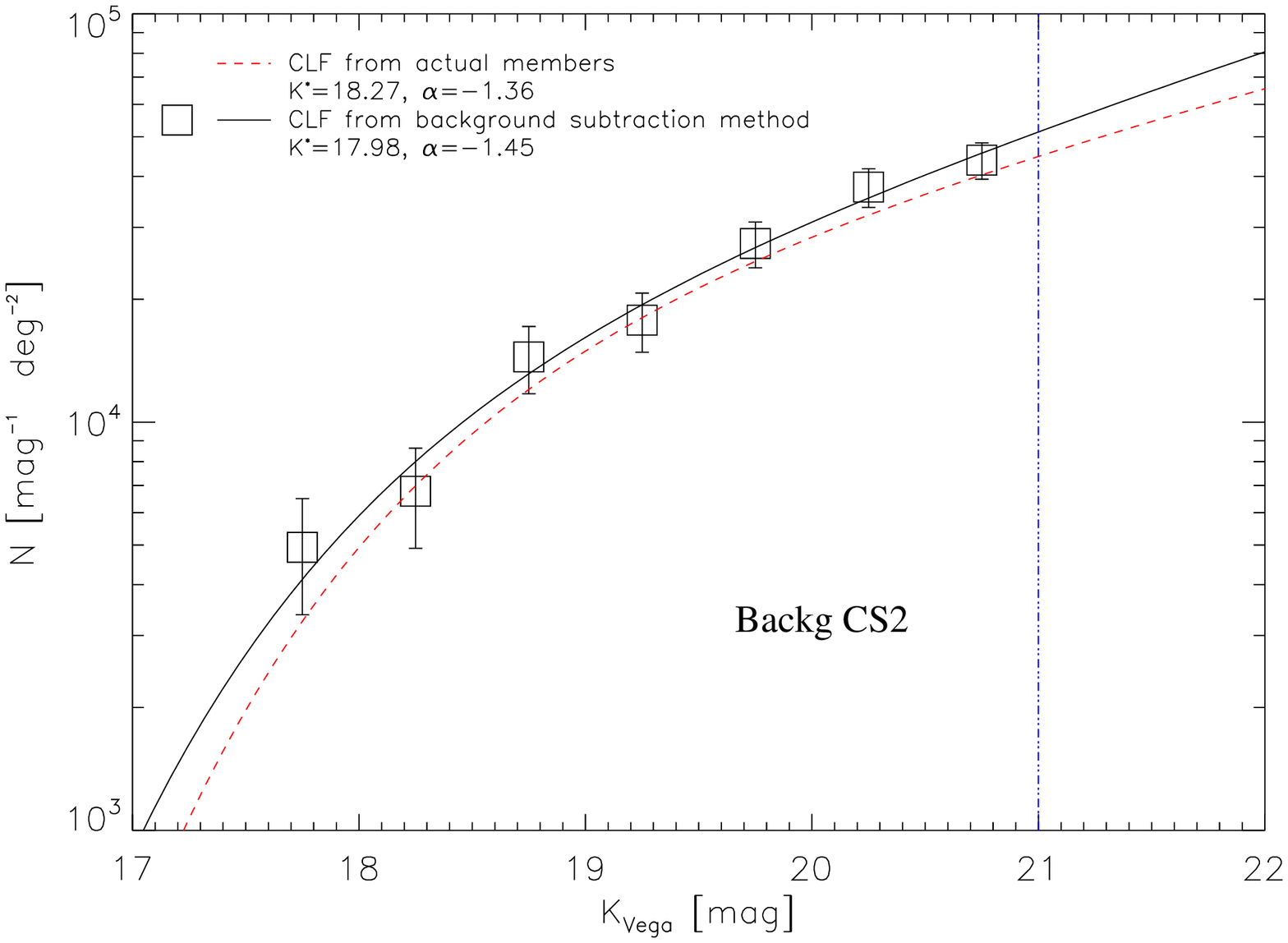}
\end{minipage}
\begin{minipage}[b]{0.5\textwidth}
\centering
\includegraphics[width=8cm]{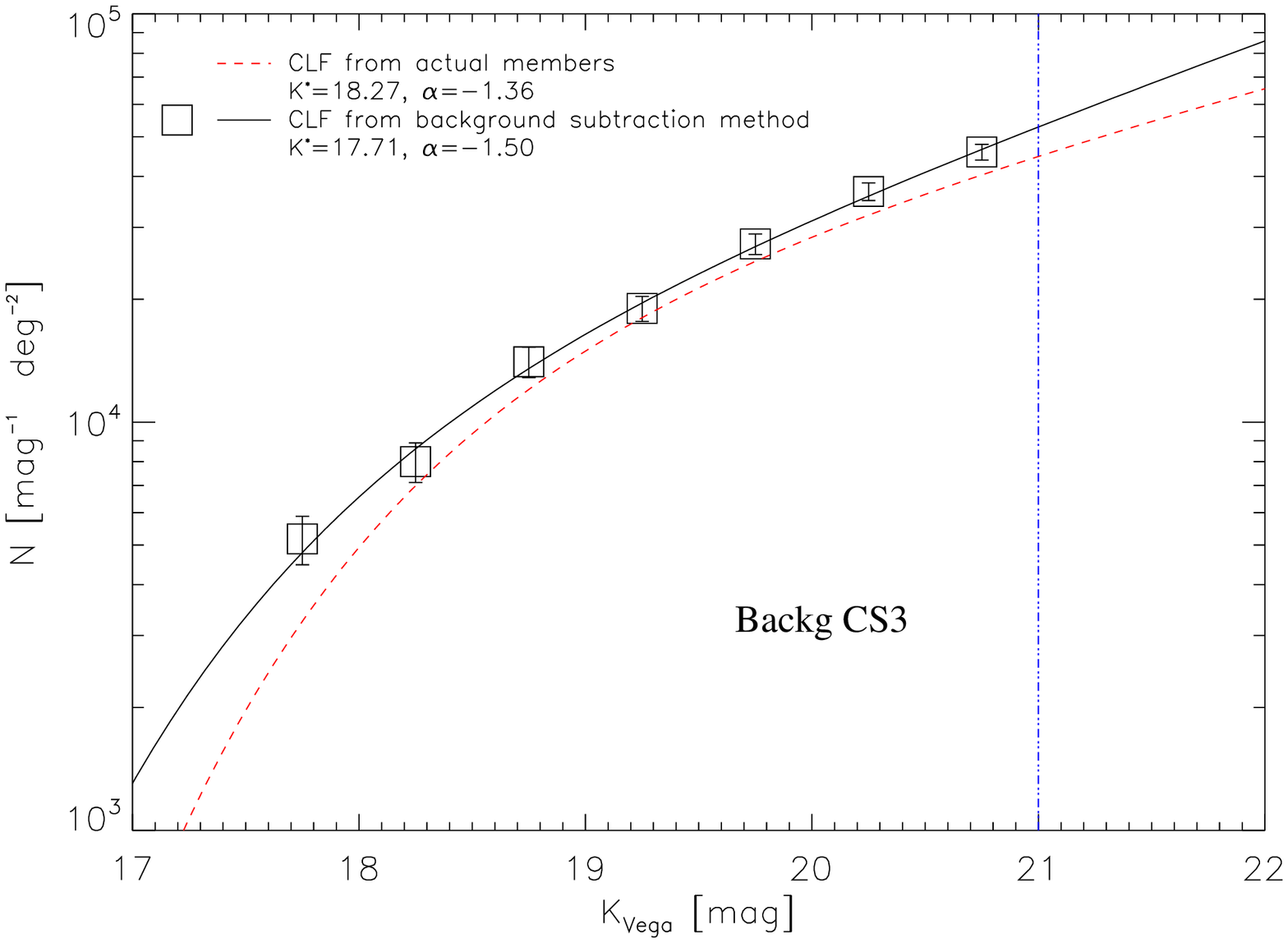}
\end{minipage}%
\begin{minipage}[b]{0.5\textwidth}
\centering
\includegraphics[width=8cm]{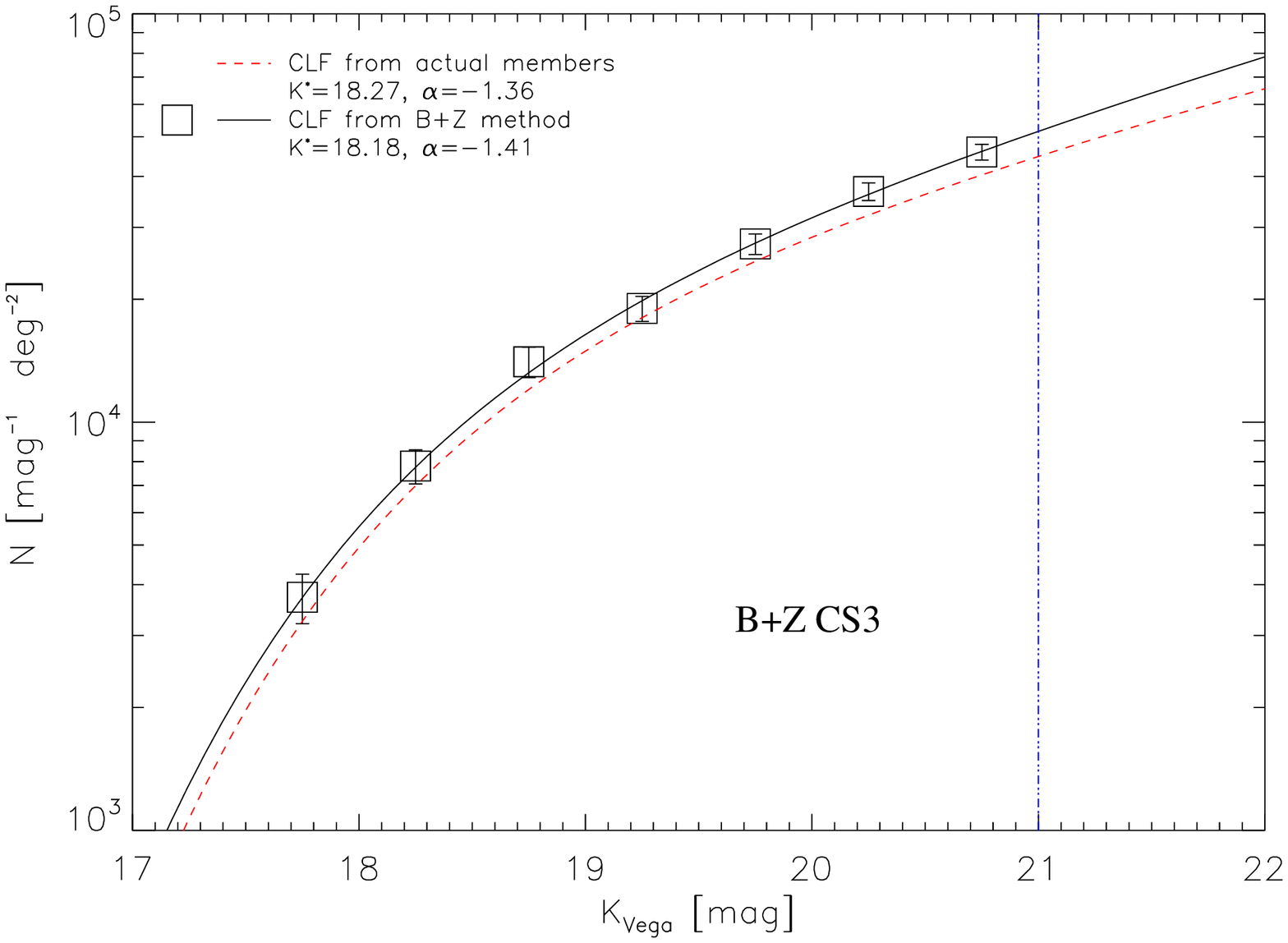}
\end{minipage}
\caption{Ks-band composite luminosity function (CLF) built from the background subtraction and B+Z methods. Shown are the CLF built from the
background subtraction method for 5 clusters (\textit{top-left}), for 10 clusters (\textit{top-right}), and for 50 clusters (\textit{bottom-left}). Also, the
CLF built from the B+Z method for 50 clusters (\textit{bottom-right}). Dashed line shows the underlying CLF, while open squares
show the recovered CLF from applying the respective method.}
\label{clf_n50}
\end{figure*}

\subsection{The composite luminosity function}

In most galaxy clusters there are too few galaxies to determine accurately the shape of the LF. A solution to this problem
is to measure the LF using a combined sample of clusters, allowing a higher precision measurement. 
This method is commonly known as the Composite Luminosity Function (CLF) method.
 
The CLF is built according to the following formulation,
\begin{equation}
N_{cj} = \frac{1}{n_j} \sum_{i=1}^{N_{clusters}} N_{ij}, 
\label{clf_equation}
\end{equation}
where $N_{cj}$ is the number of galaxies in the j-th bin of the CLF, $N_{ij}$ is the number of galaxies in the j-th 
bin of the i-th cluster, and $n_j$ is the number of clusters with limiting magnitude deeper than the j-th bin. 
The formal errors of the CLF were computed according to,
\begin{equation}
\sigma_{N_{cj}} = \frac{1}{n_j} \sqrt{\sum_{i=1}^{N_{clusters}}  \sigma_{N_{ij}}^2  }
\end{equation}
where $\sigma_{N_{ij}}$ is the formal error of the number of galaxies in the j-th bin of the i-th cluster.

For computing the CLF, we selected several galaxy clusters with masses about $1.5 \times 10^{14}\; 
\mathrm{M}_{\odot}$, located between redshifts $0.90 < z < 1.10$.
The observed CLF was built for 3 cluster samples taken from the mocks: the first sample contains 5 clusters (CS1), 
the second, 10 clusters (CS2), and the third, 50 clusters (CS3). We reset the completeness limit 
in our mock catalogues to
$\mathrm{Ks}=21.0$ in order to match the RCS nIR imaging (\citealt{mun08}), and 
we apply both, the background subtraction and B+Z methods for computing the CLF.

Figure \ref{clf_n50} shows the resulting CLF computed for a CS3 dataset,
using both the background subtraction and B+Z methods (upper and lower panels, respectively). 
The open symbols correspond to the underlying 
CLF calculated using all the cluster members, and the filled circles correspond to the CLF 
obtained using only data available in observational samples.  This figure will be discussed into
more detail in the following section.

\section{Analysis}

\subsection{Schechter parameters}

In order to study the reliability of the background subtraction and B+Z methods to recover the underlying LF, 
we fit a Schechter function to the observed CLF and then compare the best fitting observed parameters to 
those of the underlying CLF.

The best fitting parameters are found using a minimum $\chi^2$ method given by,
\begin{eqnarray}
\chi^2  & = & \sum_{j=1}^{N_{bins}} \left( \frac{N_{cj} - N(m_j;m^*;\alpha)}{\sigma_{N_{cj}}} \right)^2, \label{chi2_1} \\
N(m_j;m^*;\alpha) & = & \int_{m_j-\frac{m_{bin}}{2}}^{m_j+\frac{m_{bin}}{2}} \phi (m) dm, \label{chi2_2}
\end{eqnarray}
where $N(m_j;m^*;\alpha)$ is the predicted value of the CLF at the magnitude bin $m_j$, $m_{bin}$ is the 
magnitude bin size, and $\phi(m)$ is the Schechter function. 

We search the global minimum in the full $\phi^* - m^* - \alpha$ space sampled by a parameter grid.
The best-fitting parameters after marginalizing over $\phi^*$,
for $\mathrm{Ks}^*$ and $\alpha$ taken as free parameters,
are summarized in the Table \ref{clf_param_free} for the three samples of mock galaxy clusters defined
in the previous Section. The systematic errors are defined as the difference between the actual and estimated value, while the stochastic errors are defined by the 68 percent confidence levels.

\begin{table}
 \centering
 \begin{minipage}{80mm}
  \caption{Best-fitting parameters for $\mathrm{Ks}^*$ and $\alpha$ as free parameters.}
   \label{clf_param_free}
  \begin{tabular}{@{}llrrr@{}}
  \hline
  Dataset & Param & Actual & \multicolumn{1}{c}{Background} & \multicolumn{1}{c}{B+Z} \\
    & & & \multicolumn{1}{c}{method} & \multicolumn{1}{c}{method} \\
  \hline
   CS1 & $K^*$ & 18.27 & $18.95^{+0.90}_{-1.23}$ & $18.94^{+0.76}_{-0.86}$ \\
   	 & $\alpha$ & -1.36 & $-1.11^{+0.61}_{-0.46}$ & $-1.10^{+0.55}_{-0.41}$ \\
  \hline
  CS2 & $K^*$ & 18.27 & $17.98^{+0.80}_{-2.27}$ & $18.56^{+0.46}_{-0.67}$ \\
  	& $\alpha$ & -1.36 & $-1.45^{+0.30}_{-0.28}$ & $-1.29^{+0.28}_{-0.27}$ \\
  \hline
  CS3 & $K^*$ & 18.27 & $17.71^{+0.46}_{-0.71}$ & $18.18^{+0.31}_{-0.39}$ \\
  	& $\alpha$ &  -1.36 & $-1.50^{+0.13}_{-0.12}$ & $-1.41^{+0.12}_{-0.12}$ \\
  \hline
  \end{tabular}
  \end{minipage}
\end{table}

\begin{table}
 \centering
 \begin{minipage}{80mm}
  \caption{Best-fitting parameters derived with a fixed value of $\alpha = -1.36$}
   \label{clf_param_fixed}
  \begin{tabular}{@{}llrrr@{}}
  \hline
  Dataset & Param & Actual & \multicolumn{1}{c}{Background} & \multicolumn{1}{c}{B+Z} \\
    & & & \multicolumn{1}{c}{method} & \multicolumn{1}{c}{method} \\
  \hline
  CS1 & $K^*$ & 18.27 & $18.40^{+0.33}_{-0.39}$ & $18.50^{+0.24}_{-0.28} $ \\  
  \hline
  CS2 & $K^*$ & 18.27 & $18.30^{+0.17}_{-0.19}$ & $18.42^{+0.15}_{-0.13}$ \\   
  \hline
  CS3 & $K^*$ & 18.27 & $18.20^{+0.09}_{-0.09}$ & $18.30^{+0.08}_{-0.06}$ \\  
  \hline
  \end{tabular}
  \end{minipage}
\end{table}

Figure \ref{confidence_clf_n50} shows the 1- and 2-$\sigma$ likelihood contours of the Schechter function parameters 
for our sample CS3 using both, the background subtraction and B+Z methods. The plus symbols mark the actual CLF 
parameters and the crosses mark the best fitting observed CLF parameters. From this figure, we can conclude that the 
B+Z method gives more accurate and better constrained results than the background subtraction method, since 
the stochastic error of $\mathrm{Ks}^*$ is reduced by almost a factor or $2$.

\begin{figure}
\includegraphics[width=7.5cm]{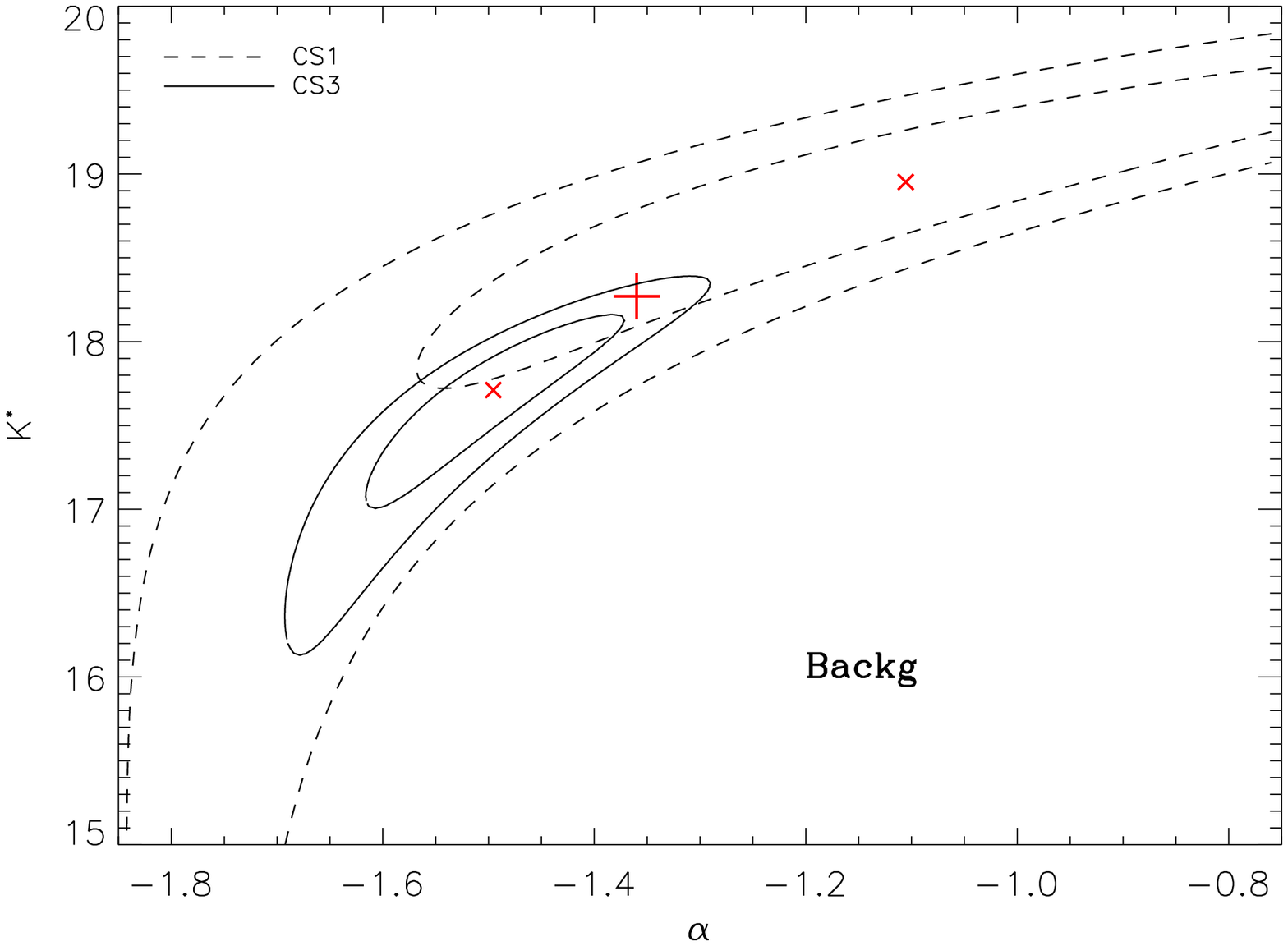}
\includegraphics[width=7.5cm]{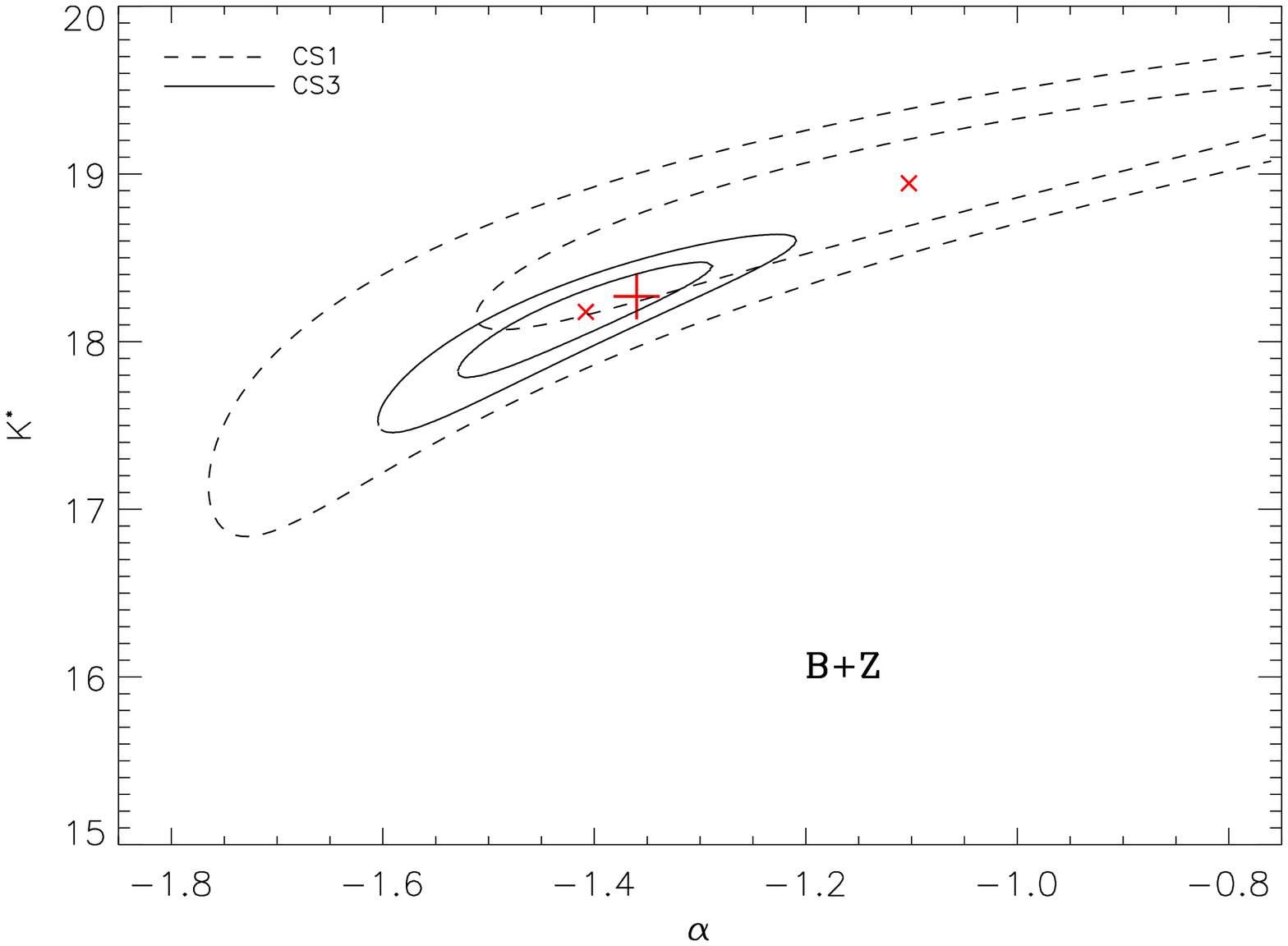}
\caption{Contour plots showing the constraints on the Schechter function parameters derived from the $\chi^2$ 
minimization analysis applied to the CLF obtained using 5 clusters (\textit{dashed line}) and 50 clusters (\textit{solid line}). 
The upper panel shows the results from applying the statistical background subtraction method; 
the lower panel shows the results from applying the B+Z method. 
The plus sign marks the parameters of the true underlying CLF, and the cross marks the best fitting parameters 
of the measured CLF. The labeled contours represent the 1- and 2-$\sigma$ confidence levels.}
\label{confidence_clf_n50}
\end{figure}

\begin{figure}
\includegraphics[width=8cm]{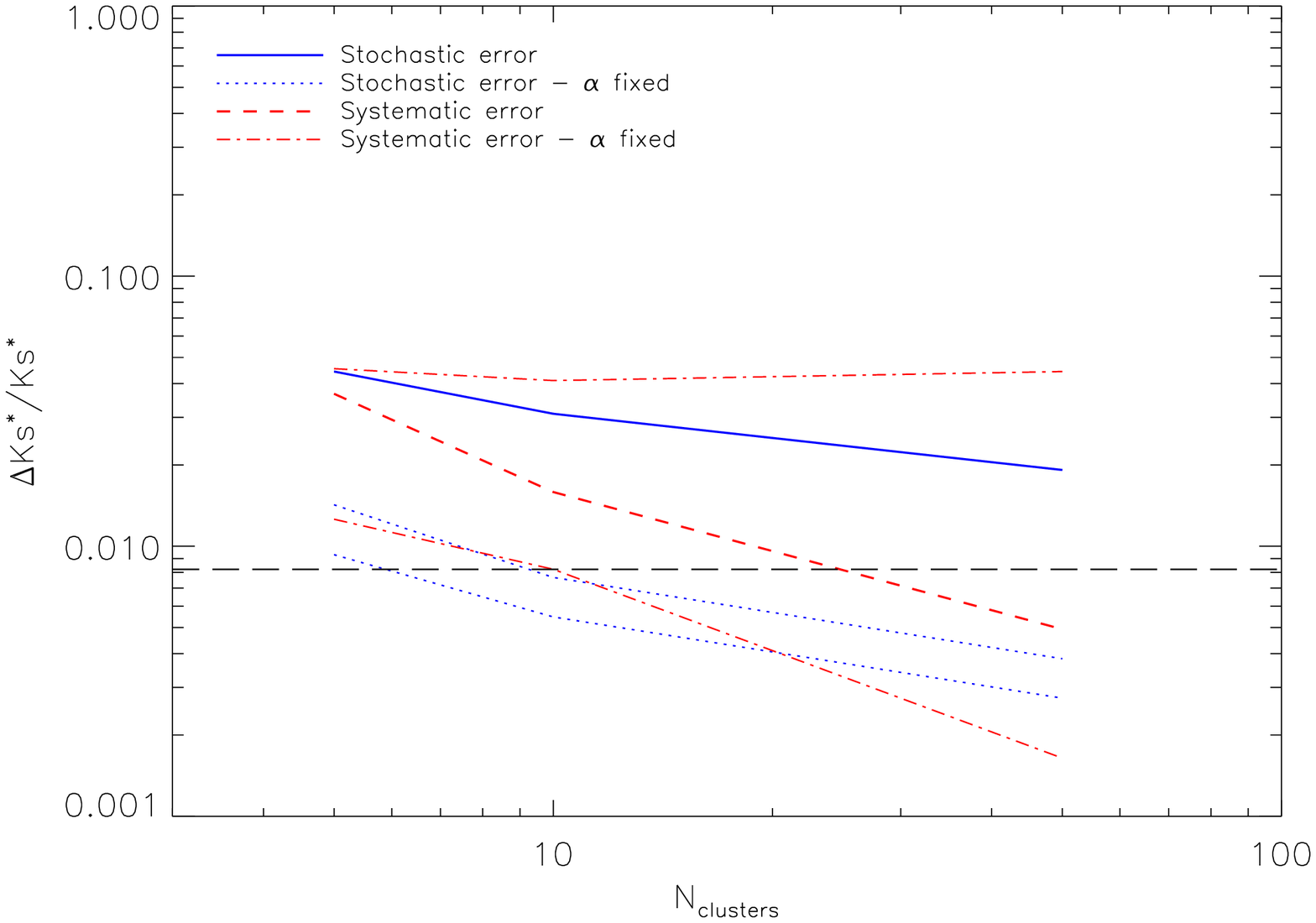}
\includegraphics[width=8cm]{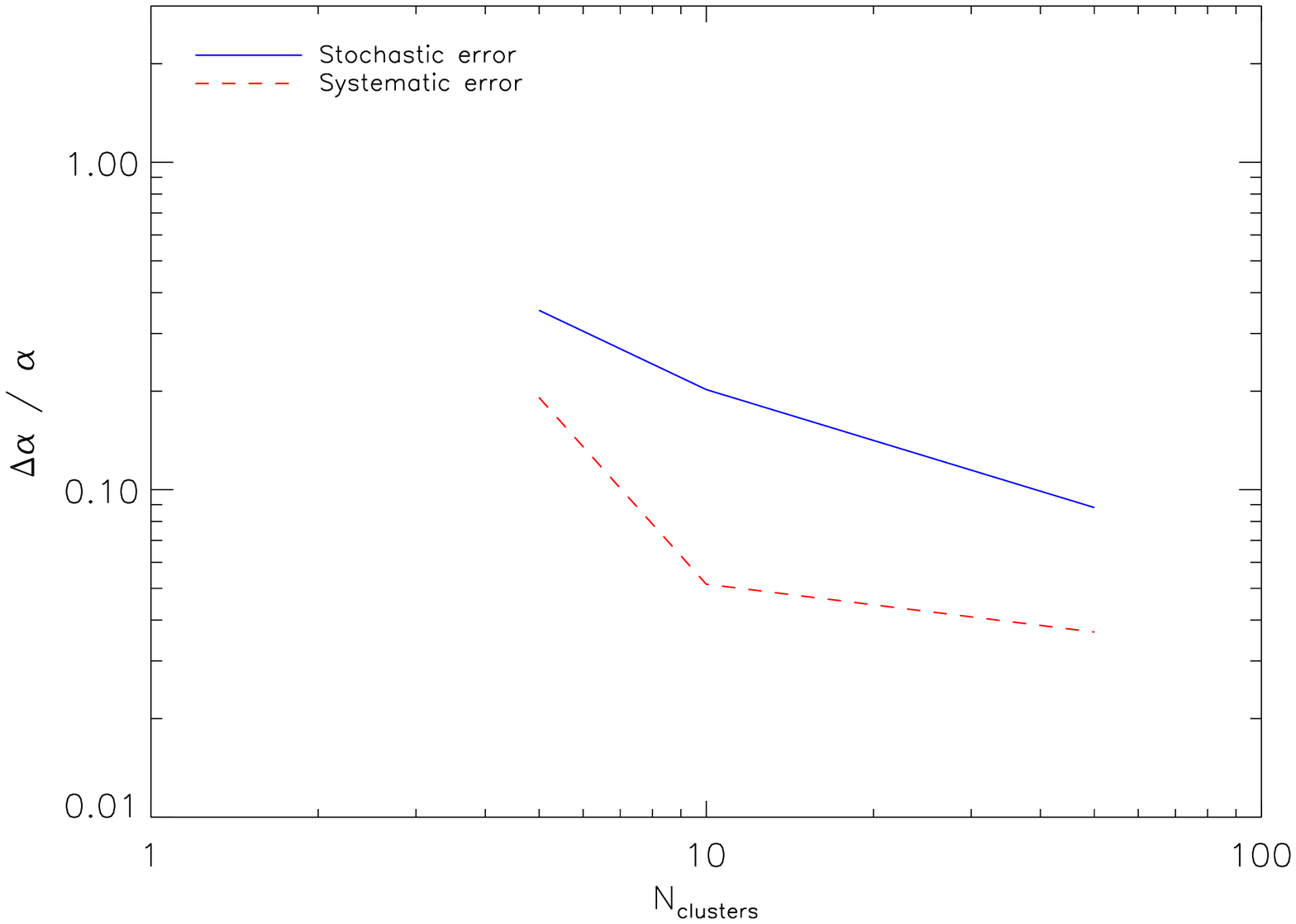}
\caption{
Upper panel: Stochastic and systematic errors in $\mathrm{Ks}^*$ obtained through the B+Z LF measurement method, for samples composed by different number of galaxy clusters, for a fixed or free value of the $\alpha$ parameter (line types are shown in the figure key). 
Upper and lower dotted lines show the stochastic errors in $\mathrm{Ks}^*$ for fixed values of $\alpha=-1.36$ and 
$\alpha=-1.00$ respectively. Upper and lower dot-dashed lines are the systematic error in 
$\mathrm{Ks}^*$ for $\alpha=-1.00$ and $-1.36$, respectively. The horizontal long-dashed line is the maximum 
error that still allows to distinguish between passive evolution scenarios with formation redshifts of $z_f=2$ and $z_f=5$.
Lower panel: Stochastic and systematic errors in $\alpha$.}
\label{clf_param_error}
\end{figure}

\begin{figure}
\includegraphics[width=7.5cm]{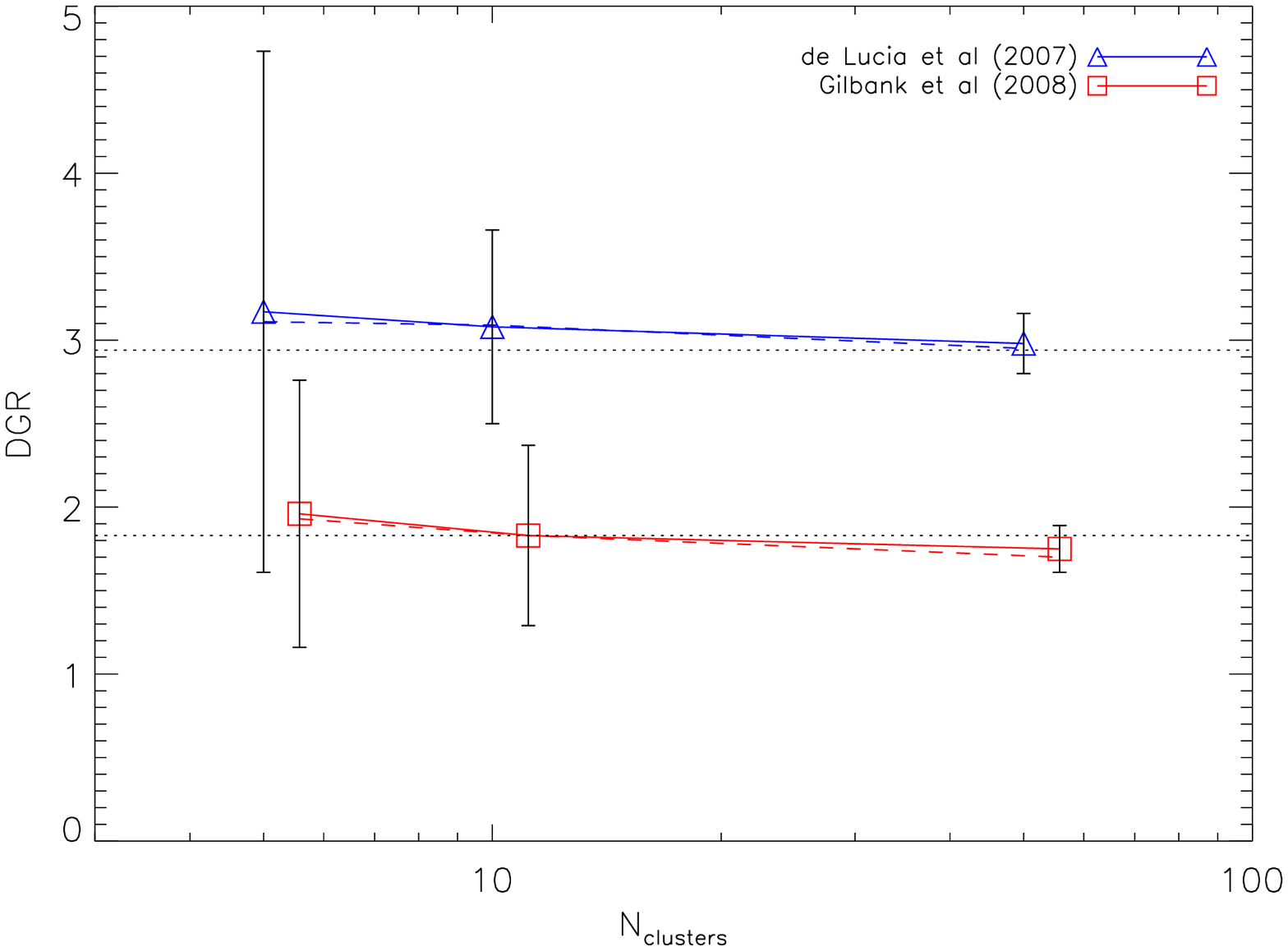}
\caption{Dwarf-to-giant ratio (DGR) for red-sequence cluster galaxies computed from background subtraction (\textit{solid line}) and photometric
redshift (\textit{dashed line}) methods. The triangles and squares correspond to 
the DGR estimated from applying the magnitude limits by \citet{del07} and \citet{gil08b}, respectively.
The underlying value is shown by the dotted line. The error-bars show the stochastic errors. 
A small horizontal offset was applied to the \citet{gil08b} results to improve clarity.}
\label{dgr_error}
\end{figure}

It is useful to study the reliability of the best LF measurement method (B+Z estimator) 
as a function of the number of galaxy clusters used 
for computing the CLF, since this can have a direct impact on current and future observational campaigns. 
In principle, a larger sample of clusters should put better constraints on the values of $\mathrm{Ks}^*$ and 
$\alpha$ as well as lower systematic errors.  Figure \ref{clf_param_error} shows the stochastic and systematic 
errors for both $\mathrm{Ks}^*$ and $\alpha$. From this figure we can conclude that the method reliability and accuracy depend
strongly on the number of galaxy clusters, as increasing the sample size by a factor of $10$ reduces the stochastic
and systematic errors in $\mathrm{Ks}^*$ by factors of $3$ and $30$, respectively.

When the parameter $\alpha$ is fixed to the true underlying value,
errors also improve, for instance, by a factor of $3$ in the characteristic luminosity. The results for the three samples of mock galaxy clusters are summarized in Table \ref{clf_param_fixed}.
In the case where the value of $\alpha$ is unknown and a fiducial value $\alpha=-1$ is assumed (as
may be the case in an observational estimate of the CLF), the systematic errors increase
significantly due to the likelihood degeneracy in the $\mathrm{Ks}^*$-$\alpha$ plane, which in
our mocks forces the resulting
characteristic luminosity to a fainter value.  It is also noticeable that in this case, the stochastic
error is lower than when fixing $\alpha$ to the true underlying values; this indicates that even though
the uncertainties in $\mathrm{Ks}^*$ diminish when fixing $\alpha$, if this parameter is offset by only a $23\%$
of its value,
there could be significant systematic errors in $\mathrm{Ks}^*$ producing biased analyses.

Regarding the faint-end slope, $\alpha$,
the stochastic and systematic
errors also diminish by about a factor of $4$ when increasing the sample from $5$ to $50$ clusters.  
It is important to notice that regardless of the number of clusters in the sample, 
the stochastic error is always found to be larger than systematic deviations.

\subsection{Mean formation epoch}

The study of the evolution of the CLF is a powerful tool to improve our knowledge on 
the processes of galaxy formation and evolution, in particular in high density regions of the Universe where
the probability of galaxy interactions with their environment is at its highest.   
In addition, these studies are easier to interpret when using infrared luminosities, since these 
are better suited than optical photometry to measure the LF, since the former 
reflect the total stellar mass of galaxies and do not depend strongly on the details of their stellar 
populations \citep*{gav96}.

One of the most interesting quantities that can be inferred from the evolution of the LF is the mean
redshift of the bulk of star formation in cluster galaxies, $z_f$; this redshift is defined as a measurement
of the time when these galaxies have already acquired most
of their stellar mass.   This is most easily obtained from observational estimates of the evolution of 
$\mathrm{Ks}^*$ by direct comparison to predictions using passive evolution models with different $z_f$.

We employ the passive evolution models of \citet{kod97} to compute the minimum number of galaxy 
clusters needed to distinguish between a scenario with $z_f=2$ and $z_f=5$ for the galaxies in our
mock catalogues.  According to these models, for a $\mathrm{L}^*$ galaxy observed at $z=1$, the difference in $\mathrm{Ks}^*$ between both scenarios is about 0.3 mag.
Therefore, in order
to compute $z_f$ with an accuracy of $2\; \mathrm{Gyr}$ at the 68 percent confidence level, the maximum systematic and stochastic errors allowed on $\mathrm{Ks}^*$ 
would be $\Delta \mathrm{Ks}^* = 0.15$ (shown as
a horizontal long-dashed line in figure \ref{clf_param_error}).
According to the results showed in figure \ref{clf_param_error} for the B+Z method, leaving 
both $\mathrm{Ks}^*$ and $\alpha$ 
as free parameters, this would require a sample with a total number of 520 clusters at $z\sim1$.  This number can be lowered
by fixing $\alpha$ to its actual value $-1.36$, as shown by the upper dotted line in figure \ref{clf_param_error}, 
since this would then require only 10 clusters in total. However, fixing $\alpha$ to the flatter value $\alpha=-1.00$, 
as shown by the lower dotted line in figure \ref{clf_param_error},
would not allow to compute an accurate value of $z_f$ and a comparison between formation scenarios would
be biased.  Another option could be to combine the measurement of ${\rm Ks}^*$ from an available observational sample, with
alternative methods to measure the formation epoch of galaxies such as Bruzual \& Charlot (2003) fits to the SEDs of
confirmed members of $z=1$ galaxy clusters, with possible restrictions from photometric data depth and number of
galaxies with such information.

\subsection{Dwarf-to-Giant ratio}

The dwarf-to-giant ratio (DGR) of red-sequence galaxies is a commonly used parameter that allows an interpretation of the
cluster galaxy population at a given redshift, providing the relative importance of the dwarf and giant galaxy population
in clusters. This parameter is simply defined as the ratio between the number of faint and bright red-sequence cluster 
galaxies within predefined magnitude ranges.  
As such, it does not require an analytic fit to the measured LF and therefore does not need
forcing any particular function to the real distribution of luminosities, while also avoiding dealing with 
parameter degeneracies.  However, it does require to define arbitrary luminosity limits for the dwarf
and giant populations that usually depend on the available data. A good review about cosmic evolution of the
DGR was published by \citet{gil08a}.

Since we are interested in studying the accuracy and bias of the DGR value computed from applying the background subtraction method},
we compute the actual and estimated DGR values from the mock catalogue.
We employ the magnitude limits published by \citet{del07}, who studied the red-sequence galaxy LF and defined 
as luminous galaxies those satisfying $M_V \le -20.0$, and as faint galaxies, $-20.0 < M_V \le -18.2$ (hereafter DEL07). 
For comparison, we also use the magnitude limits proposed by \citet{gil08b}, who defined luminous galaxies as those
within the range $-22.7 < M_V \le -20.7$, and faint galaxies those within $-20.7 < M_V \le -19.7$ (hereafter GIL08).
We convert these rest-frame V-band magnitudes to our observer-frame Ks-band, using the same method as \citet{del07}.
We use the GALAXEV stellar population synthesis code \citep{bru03} to generate model galaxy SEDs arising from a single burst stellar population formed at $z_f=3$.

In order to compute the actual value of the DGR, we build the CLF for cluster galaxies in the red-sequence from 
100 mock clusters. In the \citet{bau05} model, the $z=1$ rest-frame color-magnitude relation (CMR)
is defined by the mean color $({\rm r-z})_{rest}= 2.89 -0.025\times{\rm z_{rest}}$ and intrinsic dispersion 
$\Delta ({\rm r-z})=0.13$.  We obtain values of $\rm{DGR_{DEL}}=2.94$ and $\rm{DGR_{GIL}}=1.83$ for the magnitude limits defined by 
DEL07 and GIL08, respectively.

The observational estimates found by DEL07 and GIL08 correspond to $DGR=1.05\pm 0.16$ at $z\sim0.75$
and $DGR=0.84\pm0.05$ at $z\sim 0.9$, respectively; 
as can be seen, the semi-analytic model by \citet{bau05} tends to produce a larger number of dwarf galaxies at a redshift of $z\sim 1$,
and given the observed evolution of DGR, the discrepancy between the theory and observations will be even higher at this redshift.
Using the parameterization of the evolution of the observed DGR presented by \citet{gil08a}, the expected value of DGR at $z=1$ for
the magnitude limits defined by DEL07 is about 0.88, which is a factor 3.3 smaller than the value computed from our mock catalogues.
Notice that the value of the faint-end slope in the model, $\alpha=-1.36$, is probably responsible for
this discrepancy, since observed values tend to favour $\alpha>-1$, although with large uncertainties. 

We use the underlying DGR values to study the systematic and stochastic errors in the DGR as a 
function of the number of clusters in the samples studied.
Figure \ref{dgr_error} shows these results, where triangles and squares were computed using the DEL07 and GIL08 magnitude limits,
respectively. The solid lines correspond to the DGR computed from the background subtraction method, while the dashed lines were computed from the photometric redshift method as explained in section 3.1 It can be seen that  systematic errors in both DEL07 and GIL08 are
small compared to the dispersion in the measurements, and also, that stochastic errors show a clear improvement as the number of clusters
increases. For the background subtraction method, the latter goes from about a $35\%$ percent error for samples of $5$ clusters to 
about a $5\%$ percent error for samples with $50$ clusters.

\subsection{Occupation numbers}

A commonly used quantity in the study of galaxy populations in clusters is the occupation number, or the
total number of galaxies above a lower limit in luminosity in each galaxy cluster of a given mass.  This
can obtained from the analysis of galaxy luminosity and correlation functions in a formalism called the Halo Model (see for instance \citealt{coo05}, \citealt{coo02}).  
A study of the correlation function of SDSS galaxies by \citet{zeh04}
shows that $z\simeq0$ clusters of galaxies of mass $1.5\times 10^{14}$h$^{-1}$Mpc 
would contain an average of
$\simeq 3$ galaxies brighter than $M_r=-21$, equivalent to $M_r^*-0.5$ \citep{bla03}.
The same calculation performed on the $z=0$ galaxy population in the \citet{bau05}
model indicates that clusters of the same mass in the model are populated by an average of $2.2\pm0.1$
galaxies, in reasonable agreement with the observational estimates.

Estimates of the LF of clusters of a given mass can be used to estimate this number at any given redshift.
In our study we use the estimated counts down to $\mathrm{Ks}^*-0.5$ to find the number of galaxies residing in
$z\simeq1$, simulated RCS clusters.  The resulting numbers depend on the number of clusters available
in the sample as is shown in Table \ref{table:occ}.
As can be seen, the measured occupation numbers are in agreement with the underlying values for CS1 and CS2
samples, although the measurement errors diminish significantly as the number
of clusters in the sample increases.  The underlying value of $0.61$ galaxies per cluster does not change
significantly when calculating this quantity using a different photometric band, since the lower
luminosity limit depends on the characteristic luminosity of the CLF; therefore, the occupation
number of simulated clusters of $M\simeq1.5\times 10^{14}$h$^{-1}$M$_{\odot}$, 
increases by a factor of $\simeq 3.6$ from $z=1$ to $z=0$.

\begin{table}
 \centering
 \begin{minipage}{80mm}
  \caption{Cluster occupation numbers derived from the measured CLFs.}
   \label{table:occ}
  \begin{tabular}{@{}lrr@{}}
  \hline
  Dataset & Measured value & Actual value\\
  \hline
  CS1 & $0.80\pm 0.42$ & $0.61$\\
  CS2 &$0.73\pm 0.30$ &$0.61$\\
  CS3 &$0.76\pm 0.07$ & $0.61$\\
  \hline
  \end{tabular}
  \end{minipage}
\end{table}

\section{Conclusions}

We studied the reliability of different LF measurement methods appropriate for the study of high redshift, $z\simeq1$,
clusters of galaxies.  We do this by using mock cluster catalogues constructed using GALFORM (\citealt{bau05}) 
semi-analytic galaxies that populate a numerical simulation of a $\Lambda$CDM Cosmology.  Our mock catalogues
mimic the RCS clusters and the GOODS/ISAAC survey, which are the targets of an ongoing observational project that
will complement RCS cluster photometry with Ks-band data, and will use the GOODS/ISAAC survey to obtain the
background contamination in cluster regions (\citealt{mun08}).  We find that the best way to characterize the
error in the background counts is by means of the jacknife method for at least $4$ jacknife subsamples, which
takes into account the varying structures in the line of sight of pencil-beam surveys.

Our studies indicate that the joint use of
photometric redshifts of bright galaxies plus background-corrected counts of faint galaxies (B+Z method), provides the
best recovery of the underlying CLF.  In the process we also found that the classical Poisson method provides
accurate errors for the galaxy counts in the
cluster direction.  The error in the best-fitting Schechter parameter $\mathrm{Ks}^*$ from B+Z method,
decreases by almost a factor of $2$ when compared to the measurements obtained from using a method
relying on background-corrected counts alone.

Even though the B+Z method provides the best results, the measured CLF shows large stochastic
errors for datasets with small numbers of galaxy clusters. For 
a dataset of 5 galaxy clusters we find 4\% and 44\% stochastic errors in $\mathrm{Ks}^*$ and $\alpha$, respectively
($\Delta \mathrm{Ks}^*\simeq 0.7$ magnitudes, and $\Delta \alpha\simeq0.26$).
However, increasing the sample size to $50$ clusters improves this result in a dramatic way, reducing these errors
to a $2\%$ and $\simeq 9\%$ 
($\Delta \mathrm{Ks}^*\simeq0.1$ magnitudes, and $\Delta \alpha\simeq0.05$).
The use of a fixed value of the faint-end slope $\alpha$ for the Schechter fit
improves the stochastic errors by a factor of $3$; systematic errors do not change significantly,
unless the assumed value of $\alpha$ is offset from the underlying value.

The accuracy of the B+Z method in recovering the underlying dwarf-to-giant ratio (DGR) depends strongly 
on the luminosity limits used to define the bright and faint galaxy population, but in general the systematic
error does not improve significantly when using larger samples.  However, there is a clear tendency to obtain smaller
stochastic errors in this quantity.  On the other hand, in order to distinguish between formation redshifts
of $z_f=2$ and $5$ (an uncertainty of $\sim2$Gyr) 
for the bright cluster members, the sample of clusters needs to contain a minimum of $\sim 10$
clusters; this result corresponds to the B+Z method for measuring the CLF, plus leaving only the characteristic
luminosity $\mathrm{Ks}^*$ as a free parameter and fixing $\alpha$ to an appropriate value.
The occupation number of clusters in the sample, that can be associated to the median cluster mass, can 
be accurately obtained when using $10$ or more clusters in total.  The accuracy of this measurement increases
with the sample size reaching a stochastic error $\simeq 3\%$ for a total of $50$ clusters.  Such a measurement
can also be used to study the evolution of the halo occupation number as a function of redshift, which
can help constrain galaxy evolution models.

In conclusion, the clear advantage of near infrared wavelengths for the study of galaxy luminosities in clusters
at relatively $z\sim1$ can be exploited to a maximum if, i) multi-band photometry allowing
a photometric redshift estimate for the brightest cluster members is available, so that
a joint method using background subtraction and photometric redshifts can be applied to calculate
the CLF, ii) a reasonable number of $> 50$ clusters is included in the sample, and iii) the available 
measurements of CLFs at these redshifts allows to assume a fixed value for the faint-end slope of
the luminosity function $\alpha$.

Regarding the modeling of the evolution of galaxies in the semi-analytic model, the current
precision in the CLF measurements does not allow us to reach firm conclusions on whether the model succeeds
to reproduce the observed galaxy population.  There is, 
though, an indication of an excess of dwarf
galaxies by approximately a factor of 3 in the model 
with respect to observations.  However, we have demonstrated that 
only a factor of $\sim 10$ increase
in sample size with respect to those used in currently available $Ks$-band CLF measurements, would make a more stringent
and decisive test possible, allowing the use of cluster galaxies at $z\sim1$ to further our understanding
of how galaxies form and evolve in the Universe.

\section*{Acknowledgments}

We thank Michael Gladders and David Gilbank for helpful discussions.

RPM acknowledges support from a Conicyt Doctoral fellowship. 
NDP was supported by a Proyecto Fondecyt Regular No. 1071006.
LFB was supported by a Proyecto Fondecyt Regular No. 1085286.
 This work benefited from support from 
``Centro de Astrof\'\i sica FONDAP" at Universidad Cat\'olica de Chile.

\bsp

\label{lastpage}


\begin{thebibliography}{99}

\bibitem[\protect\citeauthoryear{Andreon, Punzi \& Grado}{Andreon et al.}{2005}]{and05} Andreon S., Punzi G., Grado A., 2005, MNRAS, 360, 727
\bibitem[\protect\citeauthoryear{Barkhouse et al.}{2006}]{bar06} Barkhouse W.A. et al., 2006, ApJ, 645, 955
\bibitem[\protect\citeauthoryear{Baugh et al.}{2005}]{bau05} Baugh C. M., Lacey C. G., Frenk C. S., Granato G. L., Silva L., Bressan A., Benson A. J., Cole S., 2005, MNRAS, 356, 1191.
\bibitem[\protect\citeauthoryear{Blanton et al.}{2003}]{bla03} Blanton M. et al., 2003, ApJ, 592, 819.
\bibitem[\protect\citeauthoryear{Bower et al.}{2006}]{bow06} Bower R. G., Benson A. J., Malbon R., Helly  J. C., Frenk C. S., Baugh C. M., Cole S., Lacey C. G., 2006, MNRAS, 370, 645.
\bibitem[\protect\citeauthoryear{Bruzual \& Charlot}{2003}]{bru03} Bruzual G., Charlot S., 2003, MNRAS, 344, 1000
\bibitem[\protect\citeauthoryear{Carlstrom et al.}{2002}]{car02} Carlstrom J. E., Holder G. P., Reese E. D., 2002, ARA\&A, 40, 643
\bibitem[\protect\citeauthoryear{Charlot}{1996}]{cha96} Charlot S., 1996, in The Universe at High-z, Large-Scale Structure, and the Cosmic Microwave Background, ed. E. Martinez-Gonzalez, J. L. Sanz (Heidelberg : Springer), 53
\bibitem[\protect\citeauthoryear{Cole et al.}{2000}]{col00} Cole S., Lacey C. G., Baugh C. M., Frenk  C. S., 2000, MNRAS, 319, 168
\bibitem[\protect\citeauthoryear{Cooray}{2005}]{coo05} Cooray A., 2005, MNRAS, 364, 303.
\bibitem[\protect\citeauthoryear{Cooray \& Sheth}{2002}]{coo02} Cooray A., Sheth R., 2002, PhR, 372, 1
\bibitem[\protect\citeauthoryear{De Lucia et al.}{2007}]{del07} De Lucia G. et al., 2007, MNRAS, 374, 809
\bibitem[\protect\citeauthoryear{De Propris et al.}{1999}]{dep99} De Propris R., Stanford S. A., Eisenhardt P. R., Dickinson M., Elston R., 1999, AJ, 118, 719
\bibitem[\protect\citeauthoryear{De Propris et al.}{2003}]{dep03} De Propris R. et al., 2003, MNRAS, 342, 725
\bibitem[\protect\citeauthoryear{De Propris et al.}{2007}]{dep07} De Propris R., Stanford S. A., Eisenhardt P. R., Holden B. P., Rosati P., 2007, AJ, 133, 2209
\bibitem[\protect\citeauthoryear{Eisenhardt et al.}{2004}]{eis04} Esisenhardt P.R. et al., 2004, ApJS, 154, 48
\bibitem[\protect\citeauthoryear{Eisenhardt et al.}{2008}]{eis08} Esisenhardt P.R. et al., 2008, astro-ph:0804.4798
\bibitem[\protect\citeauthoryear{Ellis \& Jones}{2004}]{eli04} Ellis S.C., Jones L.R., 2004, MNRAS, 348, 165
\bibitem[\protect\citeauthoryear{Gehrels}{1981}]{geh81} Gehrels N., 1986, ApJ, 303, 336
\bibitem[\protect\citeauthoryear{Garilli et al.}{1999}]{b3} Garilli B., Maccagni D., Andreon S., 1999, A\&A, 342, 408
\bibitem[\protect\citeauthoryear{Gavazzi, Pierini \& Boselli}{Gavazzi et al.}{1996}]{gav96} Gavazzi G., Pierini D., Boselli A., 1996, A\&A, 312, 397
\bibitem[\protect\citeauthoryear{Gilbank \& Balogh}{Gilbank et al.}{2008a}]{gil08a} Gilbank D., Balogh M., 2008a, astro-ph:0801.1930
\bibitem[\protect\citeauthoryear{Gilbank et al.}{2008b}]{gil08b} Gilbank D., Yee H.K.C., Ellingson E., Gladders M.D., Loh Y.-S., Barrientos L.F., Barkhouse W. A., 2008b, ApJ, 673, 742
\bibitem[\protect\citeauthoryear{Gladders \& Yee}{Gladders et al.}{2005}]{gla05} Gladders M.D., Yee H.K.C., 2005, ApJS, 157, 1
\bibitem[\protect\citeauthoryear{Gladders et al.}{2007}]{gla07} Gladders M.D., Yee H.K.C., Majumdar S., Barrientos L.F., Hoekstra H., Hall P.B., Infante L., 2007, ApJ, 655, 128
\bibitem[\protect\citeauthoryear{Gonzalez et al.}{2001}]{gon01} Gonzalez A.H., Zaritsky D., Dalcanton J.J., Nelson A., 2001, ApJS, 137, 117
\bibitem[\protect\citeauthoryear{Goto et al.}{2002}]{got02} Goto T. et al., 2002, PASJ, 54, 515
\bibitem[\protect\citeauthoryear{Iovino et al.}{2005}]{iov05} Iovino A. et al., 2005, A\&A, 442, 423
\bibitem[\protect\citeauthoryear{Kodama \& Arimoto}{1997}]{kod97} Kodama T., Arimoto N., 1997, A\&A, 320, 41
\bibitem[\protect\citeauthoryear{Kodama et al.}{2003}]{kod03} Kodama T. et al., 2003, MNRAS, 346, 1
\bibitem[\protect\citeauthoryear{Labb\'e et al.}{2003}]{lab03} Labb\'e I. et al., 2003, AJ, 125, 1107
\bibitem[\protect\citeauthoryear{Lagos, Cora \& Padilla}{2008}]{lag08} Lagos C., Cora S. \& Padilla N., 2008, accepted for publication in MNRAS, arXiv:0805.1930
\bibitem[\protect\citeauthoryear{Lin et al.}{2004}]{lin04} Lin Y.T., Mohr J., Stanford S.A., 2004, ApJ, 610
\bibitem[\protect\citeauthoryear{Madau, Pozzetti, \& Dickinson}{2004}]{mad04} Madau P., Pozzetti L., Dickinson M., 1998, ApJ, 498, 106
\bibitem[\protect\citeauthoryear{Mannucci et al.}{2001}]{man01}Mannucci F., Basile F., Poggianti B. M., Cimatti A., Daddi E., Pozzetti L., Vanzi L., 2001, MNRAS, 326, 745.
\bibitem[\protect\citeauthoryear{Marconi et al.}{2004}]{mar04} Marconi A., Risaliti G., Gilli R., Hunt L., Maiolino R., Salvati M., 2004, MNRAS, 351, 169
\bibitem[\protect\citeauthoryear{Mu\~noz et al.}{2008}]{mun08} Mu\~noz R. P., Barrientos L.F., Gladders M.D, Yee H.K.C, 2008, in preparation
\bibitem[\protect\citeauthoryear{Oemler}{1974}]{oem74} Oemler A. Jr, 1974, ApJ, 194, 1
\bibitem[\protect\citeauthoryear{Padilla \& Lambas}{Padilla et al.}{1999}]{pad99} Padilla N., Lambas D.G., 1999, MNRAS, 310, 21
\bibitem[\protect\citeauthoryear{Padilla et al.}{2004}]{pad04} Padilla N., et al. (The 2dFGRS Team), 2004, MNRAS, 352, 211.
\bibitem[\protect\citeauthoryear{Paz, Stasyszyn \& Padilla}{2008}]{paz08} Paz D., Stasyszyn F. \& Padilla N., 2008, submitted to MNRAS, arXiv:0804.4477
\bibitem[\protect\citeauthoryear{Pimbblet et al.}{2002}]{pim02} Pimbblet K., Smail I., Kodama T., Couch W., Edge A., Zabludoff A., O'Hely E., 2002, MNRAS, 331, 333
\bibitem[\protect\citeauthoryear{Postman et al.}{1996}]{pos96} Postman M., Lubin L.M., Gunn J.E., Oke J.B., Hoessel J.G., Schneider D.P., Christensen J.A., 1996, AJ, 111, 615
\bibitem[\protect\citeauthoryear{Press et al.}{1993}]{b7} Press W.H., Teukolsky S.A., Vetterling W.T., Flannery B.P., 1993, Numerical Recipies, Cambridge Univ. Press, Cambridge
\bibitem[\protect\citeauthoryear{Quadri et al.}{2006}]{b8} Quadri R. et al., 2006, AJ, submitted (astro-ph/0612612)
\bibitem[\protect\citeauthoryear{Retzlaff et al.}{2008}]{ret08} Retzlaff J. et al., 2008, in preparation
\bibitem[\protect\citeauthoryear{Romer et al.}{2001}]{rom01} Romer K., Viana P. T. P., Liddle A. R., Mann R. G., 2001, ApJ, 547, 594
\bibitem[\protect\citeauthoryear{Shankar et al.}{2004}]{shankar04} Shankar F., Salucci P., Granato G., De Zotti G., Danese L., 2004, MNRAS, 354, 1020
\bibitem[\protect\citeauthoryear{Sheth}{2007}]{she07} Sheth R., 2007, MNRAS, 378, 709
\bibitem[\protect\citeauthoryear{Springel, Frenk \& White}{2006}]{spr06} Springel V., Frenk C. S., White S. D. M., 2006, Nat, 440, 1137
\bibitem[\protect\citeauthoryear{Stanford et al.}{1997}]{sta97} Stanford S.A. et al., 1997, AJ, 114, 2232
\bibitem[\protect\citeauthoryear{Stanford et al.}{2006}]{sta06} Stanford S.A. et al., 2006, ApJ, 646, 13
\bibitem[\protect\citeauthoryear{Strazzullo et al.}{2006}]{str06} Strazzullo V., Rosati P., Stanford S. A., Lidman C., Nonino M., 
Demarco R., Eisenhardt P. E., Ettori S., Mainieri V., Toft S., 2006, A\&A, 450, 909
\bibitem[\protect\citeauthoryear{Tanaka et al.}{2007}]{tan07} Tanaka M., Kodama T., Kajisawa M., Bower R., Demarco R., Finoguenov A., Lidman C., Rosati P., 2007, MNRAS, 377, 1206
\bibitem[\protect\citeauthoryear{Toft et al.}{2004}]{tof04} Toft S., Mainieri V., Rosati P., Lidman C., Demarco R., Nonino M., Stanford S. A., 2004, A\&A, 422, 29
\bibitem[\protect\citeauthoryear{Toft, Soucail \& Hjorth}{Toft et al.}{2003}]{tof03} Toft S., Soucail G., Hjorth J., 2003, MNRAS, 344, 337
\bibitem[\protect\citeauthoryear{Valotto, Moore \& Lambas}{Valotto et al.}{2001}]{val01} Valotto C., Moore B., Lambas D., 2001, ApJ, 546, 157
\bibitem[\protect\citeauthoryear{Zehavi et al.}{2004}]{zeh04} Zehavi I., et al., 2004, ApJ, 608, 16
\bibitem[\protect\citeauthoryear{Zheng et al.}{2005}]{zhe05} Zheng Z., Berlind A. A., Weinberg D.,  Benson A., Baugh C. M., Cole S., Dav\'e R., Frenk C. S., Katz N., Lacey C. G., 2005, ApJ, 633, 791
\bibitem[\protect\citeauthoryear{Zwicky}{1957}]{zwi57} Zwicky F., 1957, Morphological Astronomy. Springer-Verlag, Berlin
\end{thebibliography}
\end{document}